%% file: main.tex
\documentclass{article}
\usepackage{arxiv}
\usepackage{lscape}

\usepackage[utf8]{inputenc} % allow utf-8 input
\usepackage[T1]{fontenc}    % use 8-bit T1 fonts
\usepackage{hyperref}       % hyperlinks
\usepackage{url}            % simple URL typesetting
\usepackage{booktabs}       % professional-quality tables
\usepackage{amsfonts}       % blackboard math symbols
\usepackage{nicefrac}       % compact symbols for 1/2, etc.
\usepackage{microtype}      % microtypography
\usepackage{lipsum}		% Can be removed after putting your text content
\usepackage{graphicx}
\usepackage{cite}
\usepackage{array}
\usepackage{amssymb, amsmath, amsthm}
\usepackage{graphicx}
\usepackage{lmodern,url}
\usepackage{makecell} % Sirve para hacer doble línea en una celda de tabla
\usepackage{cancel}
\usepackage{multirow}
\usepackage{microtype}
\usepackage{lineno}
\usepackage{xspace}
\usepackage{xcolor}
\usepackage{siunitx}
\usepackage{todonotes}
\usepackage{booktabs}
\usepackage[ruled,vlined]{algorithm2e}
\usepackage{optidef}
\usepackage{mathdots}
\usepackage{subcaption}
\usepackage{csquotes}
\usepackage{caption}
\graphicspath{{Figures/}}

\usepackage{amsmath}
\usepackage{tikz}
\usetikzlibrary{fit}
\usetikzlibrary{positioning}
\tikzset{%
  highlight/.style={rectangle,rounded corners,fill=red!15,draw,fill opacity=0.25,thick,inner sep=0pt}
}

\definecolor{mygreen}{RGB}{160, 242,182}%160, 242, 182

\definecolor{mypurple}{RGB}{236, 223, 234}

\newcommand{\RH}{\ensuremath{\mathcal{R}_H}\xspace}
\newcommand{\aH}{\ensuremath{a_{\small H}}\xspace}
\newcommand{\OutHouseRepNum}{\ensuremath{\rho_i}\xspace}
\newcommand{\avOutHouseRepNumI}{\ensuremath{\bar{R}_{\rm out}}\xspace}
\newcommand{\avOutHouseRepNumRH}{\ensuremath{R_{{\rm out},t}}\xspace}

\newcommand{\avOutHouseRepNumCrit}{\ensuremath{R_{{\rm out}}^{\rm crit}}\xspace}
\newcommand{\momratio}{\ensuremath{\eta^{*}}\xspace}%{\ensuremath{\frac{m_{2}}{E}}\xspace}
\newcommand{\momRatio}{\momratio}%{\ensuremath{\frac{m_{2}}{E}}\xspace}
\newcommand{\hhDist}{\ensuremath{\eta}\xspace}
\newcommand{\Boost}{\ensuremath{\mathcal{B}}\xspace}
\allowdisplaybreaks

\title{Household size can explain 40\% of the variance in cumulative COVID-19 incidence across Europe}

\usepackage{authblk}

\author[1,2$\dagger$*]{Seba Contreras}
\author[1,2$\dagger$]{Philipp Dönges}
\author[3$\dagger$]{Maciej Filinski}
\author[1,2$\dagger$]{Joel Wagner}
\author[3]{Viktor Bezborodov}
\author[3]{Marcin Bodych}
\author[4]{Barbara Pabjan}
\author[5]{Franciszek Rakowski}
\author[6$\ddagger$]{Jan Pablo Burgard}
\author[3$\ddagger$*]{Tyll Krueger}
\author[1,2$\ddagger$*]{Viola Priesemann}

\affil[1]{Max Planck Institute for Dynamics and Self-Organization, G\"ottingen, Germany.}
\affil[2]{Institute for the Dynamics of Complex Systems, University of G\"ottingen, G\"ottingen, Germany.}
\affil[3]{Wroclaw University of Science and Technology, Poland.}
\affil[4]{Institute of Sociology, University of Wroclaw, Poland.}
\affil[5]{Interdisciplinary Centre for Mathematical and Computational Modelling, University of Warsaw, Poland.}
\affil[6]{Trier University, Germany.}

\affil[ ]{{$*$} Corresponding Authors: Seba Contreras (seba.contreras@ds.mpg.de), Tyll Krüger (tyll.krueger@pwr.edu.pl), Viola Priesemann (viola.priesemann@ds.mpg.de)}
\affil[ ]{{$\dagger,\ddagger$} These authors contributed equally}
\date{}

\renewcommand{\headeright}{ }
\renewcommand{\undertitle}{ }

\begin{document}
%TC:ignore
\maketitle

\begin{abstract} %unstructured, 125 words

Household size impacts the spread of respiratory infectious diseases: Larger households tend to boost transmission by acquiring external infections more frequently and subsequently transmitting them back into the community. Furthermore, mandatory interventions primarily modulate contagion between households rather than within them. We developed an approach to quantify the role of household size in epidemics by separating within-household from out-household transmission, and found that household size explains 41\% of the variability in cumulative COVID-19 incidence across 34 European countries (95\% confidence interval: [15\%, 46\%]). The contribution of households to the overall dynamics can be quantified by a boost factor that increases with the effective household size, implying that countries with larger households require more stringent interventions to achieve the same levels of containment. This suggests that households constitute a structural (dis-)advantage that must be considered when designing and evaluating mitigation strategies.

\end{abstract}
%\vspace*{1cm}
%\scc{PAD with ToDos: \url{https://pad.gwdg.de/R_nuAF6oQcuYty3Hyt9VGw?edit}}
%TC:endignore
\clearpage

\section*{Introduction} 
%%%%%%%%%%%%%%%%%%%%%%

Emerging and re-emerging respiratory infectious diseases, exemplified by the recent COVID-19 pandemic, highlight the urgent need for robust pandemic preparedness plans. This requires identifying the factors that determine how a disease spreads within the population, thereby providing a basis for the informed design of public health interventions, such as non-pharmaceutical interventions (NPIs). However, these measures typically target community transmission and have little effect in controlling transmission within households \cite{tomori2021individual,wong2023social,doi:10.1137/23M1556861}. In addition, household contacts differ from external interactions: they are prolonged and more proximal, thereby increasing the likelihood of transmission. Once a pathogen enters a household, the probability of spreading to other members is high, making households potent accelerators of disease transmission \cite{van2024estimation,mogelmose2023population,bidari2024impact}. Therefore, to implement effective NPIs, it is critical to understand how household structure (i.e., household size distribution) impacts epidemics\cite{hilton2019incorporating,ball1997epidemics,ball2002epidemics,mogelmose2023population,goetz2024modeling,doenges2023sirmodel,house2008deterministic,house2009household}.

Across countries, incidence, prevalence, and excess mortality varied widely, also for the  COVID-19 pandemic. Across Europe, this variation cannot exclusively be attributed to differences in how NPIs were implemented \cite{karlinsky2021tracking,sharma_understanding_2021,hale2021global,wang2022estimating}. While differences in demographics \cite{odriscoll_age-specific_2021}, culture \cite{mogi2022influence,gokmen2021impact}, trust in the government \cite{alfano2024unlocking}, and resources \cite{fernandez2022socioeconomic} can explain some of the variation, all these factors covary with a country's most fine-grained functional unit: household structures. It thus remains open to determine how much of the observed variation in prevalence can be explained by household size alone.

To address this question, we first developed a mathematical framework to distinguish between within- and out-household disease spread. Broadly speaking, within-household contacts "boost" the spread of the disease, whereas out-household contacts modulate the overall reach of an outbreak. We then quantified the out-household spread for 34 European countries, allowing us to rank them according to their mitigation of out-household spread. Noting that households are also part of major socioeconomic indicators correlated with the observed prevalence, such as the Human Development Index (HDI), we finally quantified the extent to which household size contributes to this correlation by computing the semi-partial correlation, controlling for household effects. These combined analyses demonstrate that nations with larger households faced an inherent disadvantage in mitigating the impacts of the COVID-19 pandemic. This is because the larger the structural boost induced by households, the lower the effective out-household spread allowed for containment.

\clearpage
\section*{Results}
\subsection*{Separating out- from within-household transmission}

\begin{figure}[h]
\hspace*{-1.0 cm}
    \centering
    \includegraphics[width=180mm]{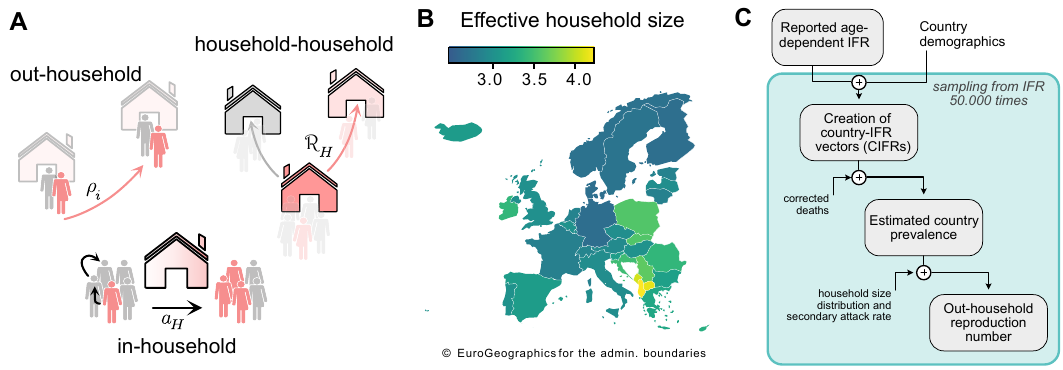}
    \caption{%
        \textbf{Disease spread from the perspective of households.} \textbf{A.} Disease spread can be categorized into the contagion occurring outside and inside of households. \OutHouseRepNum quantifies the number of potentially contagious contacts of an individual $i$. The out-household reproduction number \avOutHouseRepNumI is defined as the mean \OutHouseRepNum, averaged over all infectious individuals in an outbreak $I$. Once an infectious individual enters a household of size $k$, the in-household secondary attack rate \aH quantifies the in-household outbreak as the fraction of the $k-1$ members who are infected. The household reproduction number \RH is defined as the expected number of households a typical infectious household infects. \textbf{B.} Variation of the effective household size $\momratio$ across European countries. \textbf{C.} Methods overview: Using age-dependent infection fatality rates (IFRs) and country demographics, we obtained country-specific IFR samples (CIFRs). Combined with an estimate for the number of COVID-19 deaths, we estimated the COVID-19 prevalence (cumulative incidence) for each country. Finally, combining country-specific prevalences, household size distributions, and in-household secondary attack rates, we computed the out-household reproduction number $\avOutHouseRepNumI$. 
        }
    \label{fig:methods_merged1-2}
\end{figure}

By separating out- from within-household disease spread, we can quantify how much of the transmission can be modified by NPIs. Focusing on the out-household contacts first, we can study disease spread in a network of households where partially infectious households infect others through bridges of mutual contact (Fig.~\ref{fig:methods_merged1-2}A). We can then define a \textit{household reproduction number} \RH as the spread between households, i.e., the expected number of households that are infected by one infectious household during the outbreak. As derived in Section~\ref{sec:supplementary-S1-1-model}, \RH can be written as the product of the out-household spread and a household-dependent "boost" factor 

\begin{equation}\label{eq:RH}
    \RH = \avOutHouseRepNumRH\, \Boost \,.
\end{equation}

\avOutHouseRepNumRH is the average out-household reproduction number at a given time $t$, defined as the expected number of individuals that a typical individual infects \textit{outside} of their household, assuming that all contacts are susceptible. As \avOutHouseRepNumRH does not include immunity or its loss, it can be understood as a basic reproduction number, adjusted to exclude within-household contagion. Knowing the household size distribution \hhDist of the country (i.e., the probability $P(k)$ that a randomly selected household consists of $k$ members) and the in-household secondary attack rate of the disease $a_H$---which quantifies the expected fraction of the household that is infected in an outbreak--- the household "boost" $\Boost$ to the spreading dynamics  can be derived as 

\begin{equation}\label{eq:boost}
    \Boost =  1+\aH \left(\momRatio-1\right),
\end{equation}

where \momRatio is the effective household size, defined as the ratio between the second and first statistical moments of \hhDist, mathematically equivalent to the mean of \hhDist plus a variance-dependent correction factor (see Methods \ref{sec:supplementary-S1-1-model} for details).  Intuitively, we subtract one from \momRatio  in~\eqref{eq:boost}, because the person who brings the infection into the household cannot be infected again. Note that \Boost solely depends on \momRatio, not on the full household size distribution \hhDist. 

The critical value $\RH = 1$ delimits the transition between subcritical and supercritical disease spread. Consequently, the critical out-household reproduction number is inverse to the household boost factor, $\avOutHouseRepNumCrit = \Boost^{-1}$, and thus depends on the effective household size $\momRatio$ (Fig.\ref{fig:critical-outhousehold}). The implications of this relationship are far-reaching: countries or regions with larger effective household size $\momRatio$ reach critical disease spread at smaller values of $\avOutHouseRepNumCrit$, and thus require stronger interventions to mitigate an outbreak than countries with smaller $\momRatio$ (see Fig.~\ref{fig:methods_merged1-2}B for an overview of effective household size across Europe). With $\momRatio$ varying between $2.6$ and $4.2$ across Europe, and assuming a within-household secondary attack rate of $a_H=0.2$, the critical reproduction number $\avOutHouseRepNumCrit$ differs between $0.75$ and $0.61$, respectively.

To calculate the \avOutHouseRepNumI, defined as the average \OutHouseRepNum over all infectious individuals in an outbreak $I$, we need to know the prevalence $\alpha$, which coincides with the cumulative incidence when re-infections are not common (in the time window considered). We proceeded as described in Fig.~\ref{fig:methods_merged1-2}C: we used age-specific infection fatality rates (IFRs) and country demographics to estimate country-specific infection fatality rates (CIFRs). With these CIFRs, we estimated COVID-19 deaths, including a fraction of the surplus deaths (similar to \cite{wang2022estimating}), and COVID-19 prevalence $\alpha$ for the time period from January 1st, 2020 (before COVID-19 cases or deaths were reported in Europe),  until June 13th, 2021. This end date was selected because it marks a period when infection numbers were consistently low across all considered countries \cite{owidcoronavirus}, which is required for accurately averaging $\avOutHouseRepNumI$ over an entire wave. Combined with the household size distribution \hhDist \cite{EurostatHHdist}, we calculate \avOutHouseRepNumI. To propagate the uncertainty in the IFR estimation, we sampled from the distributions reported in \cite{brazeau2020report,brazeau2022estimating} and repeated the process \num{50000} times, obtaining distributions of prevalences and \avOutHouseRepNumI. Details and robustness checks are provided in the Supplementary Material, Sections~\ref{sec:supplementary-S1-1-model} and~\ref{sec:supplementary-S3-robustness} respectively.
 
\subsection*{Effective household size as a structural (dis-)advantage for pandemic control}

\begin{figure}[!hb]
%\hspace*{-1.0 cm}
    \centering
    \includegraphics[width=120mm]{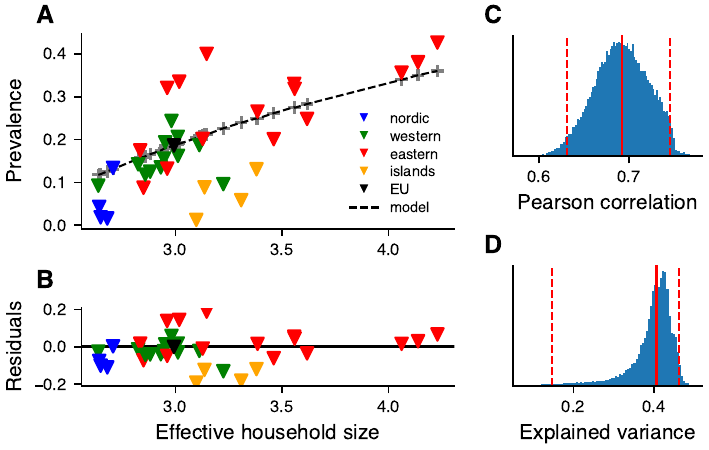}
    \caption{%
        \textbf{Differences in the effective household size can explain about 40\% of the variation in COVID-19 prevalence across European countries in the first pandemic year.} \textbf{A.} We analyzed a hypothetical scenario where all European countries had the same out-household spread (computed by treating Europe as a single country), namely, the same conditions (NPIs, climate, social culture) except for their household size distribution. By doing so, we calculated theoretical prevalences (gray crosses and black dashed line) and interpreted deviations from them as differences in the effective reduction of out-household contagion across countries (\textbf{B}). Besides the strong correlation between observed prevalence and effective household size (\textbf{C}), the latter explains 40.6\% (95\% CI: [14.7\%, 46.1\%]) of the variance observed in the former (\textbf{D}). Countries are color-coded as follows: Nordic countries (blue), Western European countries (green), and post-communist countries of Central and Eastern Europe (red), and island countries (yellow). Note that, following the method in Fig.~\ref{fig:methods_merged1-2}C, the result is a distribution of theoretical prevalence, which is not shown in this figure for clarity. A larger version with country labels is available in the Supplementary Section~\ref{fig:supplementary-mainresults_withlabels}.    
        }
    \label{fig:results_1}
\end{figure}

We derived the impact of household size on COVID-19 spread, and thus on the prevalence $\alpha$, by assuming (counterfactually) that all European countries had the same out-household reproduction number $\bar{R}_{\rm out, Europe}$. As described in the Methods section, this assumption enabled us to quantify the impact of the effective household size on $\alpha$ (gray crosses, Fig.~\ref{fig:results_1}A). Differences between the expected (gray) and observed (colored) prevalence account for variations in local epidemic management, specifically the effectiveness of mitigation measures. Thus, they indicate how well the disease was mitigated in a given country, factoring out the contribution of households. Countries below the theoretical curve achieved a higher reduction in out-household contagion compared to those above it (residuals in Fig.~\ref{fig:results_1}B). Note that this apparent success may reflect effective interventions but also differences in other factors, such as culture, compliance, resources, and weather.

As predicted from Eq.~\eqref{eq:RH}, the theoretical prevalence increases with the effective household size $\momRatio$. Furthermore, there is a clear correlation between the effective household size $\momRatio$ and the observed prevalence $\alpha$ (Pearson's correlation 0.69 (95\% CI: [0.63, 0.75]), p-value $\num{4.16e-6}$  (95\% CI: [$\num{2.78e-7}$, $\num{4.76e-5}$]) Fig.~\ref{fig:results_1}C). Quantitatively, $\momRatio$ explains 40.6\% (95\% CI: [14.7\%, 46.1\%]) of the variance in the data (Fig.~\ref{fig:results_1}D). 

\subsection*{Comparison of out-household contagion across European countries}

\begin{figure}[!ht]
\hspace*{-1.0 cm}
    \centering
    \includegraphics[width=180mm]{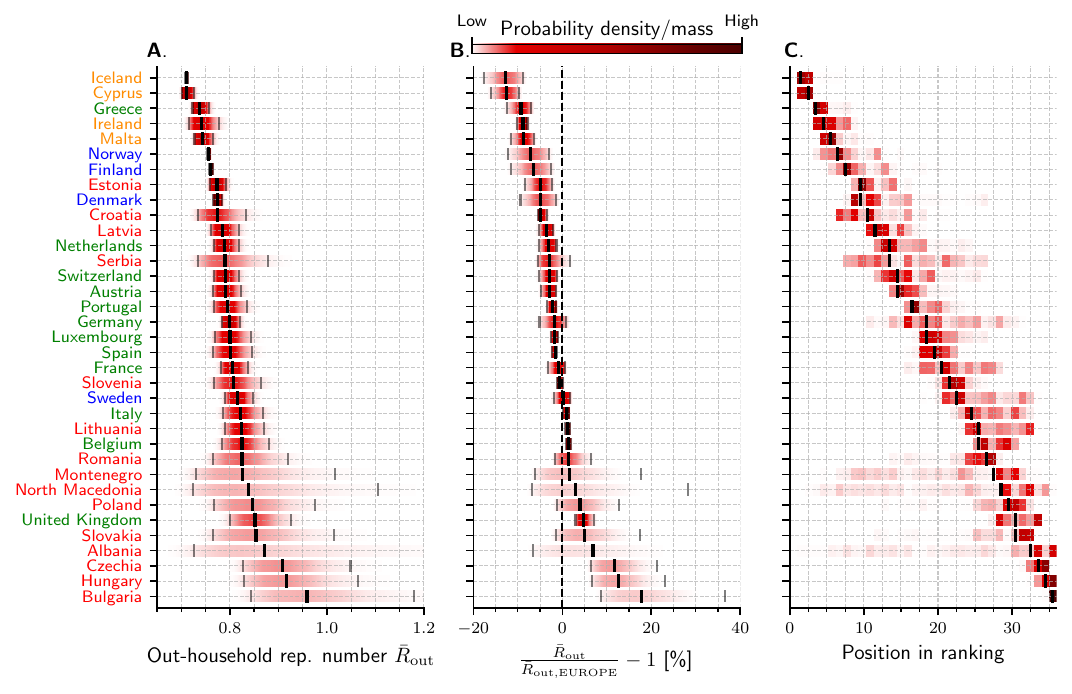}
    \caption{\textbf{Out-household COVID-19 spread across European countries.} 
    \textbf{A, B.} Using our prevalence estimations and country-specific infection fatality rates (CIFR), we computed the $\avOutHouseRepNumI$ distributions (\textbf{A}) and their deviations from the European mean (\textbf{B}) (countries ranked in order of increasing median) for the timeframe between January 1st, 2020, and June 13th, 2021. \textbf{C.} As positions in the ranking for different CIFR vectors are correlated, we obtained a distribution over the possible ranking spots for each country. Vertical lines represent the median and 95\% CIs, and color bars represent the probability density or mass.}    \label{fig:ranking}
\end{figure}

Quantifying out-household disease spread via $\avOutHouseRepNumI$ offers a way to assess the effectiveness of NPIs in each country, factoring out the contribution of households to the overall dynamics. Using the observed prevalence for each country, we obtained country-specific distributions for \avOutHouseRepNumI (Fig.~\ref{fig:ranking}A) and its deviation from the European mean (Fig.~\ref{fig:ranking}B), and sorted the countries in order of increasing median. This sorting induces a ranking for the (out-household) effectiveness of NPIs. As the positions in the ranking may vary for different sets of CIFR vectors, we obtained a distribution over the ranking positions for each country (Fig.~\ref{fig:ranking}C). As in Fig.~\ref{fig:results_1}, the same groups of countries fall into similar ranges (Fig.~\ref{fig:ranking}), with countries of Central and Eastern Europe displaying higher values of $\avOutHouseRepNumI$ and thus ranking lower than the Nordic and island countries. It is thus necessary to understand the cultural and sociological processes behind the household structure and size that may induce such clustering. 

The structure and size of households result from long-term economic and cultural processes, hence their variation across countries. For example, in 2020, there were more single-adult households without children and single-parent households with children in Nordic countries than in Central and Eastern Europe \cite{eurostat_news_2021}. Furthermore, a higher share of extended family households was found in central-eastern Europe, particularly in Bulgaria (17.3\%), Poland (11.1\%), Latvia (11.1\%), and Romania (10.3\%), whereas in Nordic countries, extended households constitute less than 1\% \cite{Eurostat2015,Eurostat2021}. As shown in Fig.~\ref{fig:ranking}A, the pattern of \avOutHouseRepNumI across countries appears to align with differences in income, with lower-income countries generally experiencing higher \avOutHouseRepNumI. In contrast, Scandinavian and Western European countries show the opposite trend. An exception lies in Island states; Greece and Cyprus, which have low incomes, managed to maintain relatively low \avOutHouseRepNumI. During the COVID-19 pandemic, Socio-geo-economic factors have been linked to the levels of morbidity and mortality \cite{elgar2020trouble,bohle2022east}, and we show that the effectiveness of interventions in reducing out-household spread also seems to follow these patterns.

\subsection*{Human development index, household size, and COVID-19 prevalence}
\label{sec:partialcorrelation}

\begin{figure}[!ht]
    \hspace*{-1cm}
    \centering    
    \includegraphics{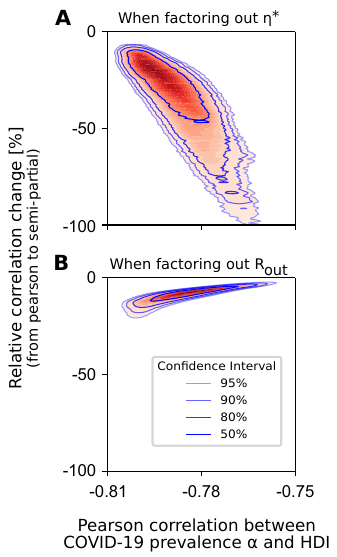}
    \caption{\textbf{Effective household size \momratio and the out-household rep. number \avOutHouseRepNumI influence the correlation between COVID-19 prevalence $\alpha$ and the Human Development Index (HDI).}  We used the semi-partial correlation to quantify the relative change in the correlation (Pearson vs. semi-partial) between $\alpha$ and HDI when removing the influence of \momratio (\textbf{A}) and \avOutHouseRepNumI (\textbf{B}) from $\alpha$.}    \label{fig:figure4-partialcorrelation}
\end{figure}    

As seen in the previous sections, households have a pronounced impact on COVID-19 prevalence. However, they are also a structural part of major socioeconomic indicators, such as the Human Development Index (HDI), that are strongly correlated with the observed prevalence $\alpha$ (median Pearson correlation coefficient between COVID-19 prevalence and HDI: $r=-0.79$, median p-value $p=\num{2.15e-8}$ Fig.~\ref{fig:figure4-partialcorrelation}). Here, using the method of semi-partial correlation, we show that the effective household size explains a substantial portion of the correlation observed between COVID-19 prevalence and HDI. The semi-partial correlation is particularly useful for this purpose, as it quantifies how much of the correlation remains between the residuals of a regression curve (a linear model when another approach is not available) and another variable.

The semi-partial correlation between $\alpha$ and the HDI after accounting for the effective household size is lower than the Pearson correlation (median $r=-0.52$, $p=0.0015$, Fig.~\ref{fig:figure4-partialcorrelation}), implying that a large portion of the correlation between prevalence and HDI can be attributed to the influence of the household distribution (median reduction of 34.4\% (95\% CI: [11.1\%, 94.3\%]) Fig.~\ref{fig:figure4-partialcorrelation}A). 
In contrast, its value when factoring out the out-household reproduction number (median $r = -0.72 $, $p = 1.09\cdot 10^{-6}$)  does not deviate as much from the original Pearson correlation, accounting for a weaker contribution (median reduction of 8.5\% (95\% CI: [3.6\%, 18.6\%]) Fig.~\ref{fig:figure4-partialcorrelation}B). 
All values and confidence intervals are available in Table~\ref{sup_tab:correlations}. This example illustrates the importance of controlling for households in correlation models and how our theory enables us to do so. 

\section*{Discussion}

Because household contacts are closer than most other social interactions, it is intuitive that they may have a substantial impact on how epidemics unfold. This work presents a conceptual and theoretical framework for quantifying the extent of this effect and provides evidence for the impact of the household size distribution on COVID-19 spread. 

Our framework enables us to isolate the contribution of households to disease spread, rewriting the epidemic threshold as the product of a household-boost factor \Boost and a critical out-household reproduction number $\avOutHouseRepNumCrit$. 
The boost factor increases linearly with the effective household size $\momRatio$, in line with the results of previous theories \cite{ball1997epidemics,ball2002epidemics,house2008deterministic}. This means that countries with larger households face a structural disadvantage and need stronger contact reduction between households to achieve the same containment levels. The out-household reproduction number \avOutHouseRepNumI factors out the contribution of households while encompassing all factors contributing to out-household disease spread, including culture, weather, overall compliance with interventions, and their effectiveness, making it a comprehensive index. We therefore propose using \avOutHouseRepNumI as a new index to measure and compare out-household contagion across countries or regions.

On the evidence side, we showed that households explain 40.6\% (95\% CI: [14.7\%, 46.1\%]) of the variance in COVID-19 incidence across 34 European countries. In fact, the correlation between the effective household size and the observed prevalence is strong (Pearson's correlation 0.69 (95\% CI: [0.63, 0.75]) and statistically significant (median p-value $\SI{4.2e-6}{}$). We identified distinct groups of countries whose observed prevalence deviates from the reference curve, forming separate clusters. Deviations from this curve provide a measure of pandemic mitigation across countries, factoring out the contribution of households.

We extensively reviewed the literature to obtain accurate parameter values for our model. Regarding the in-household secondary attack rate $a_H$, although increasing, the overall value across 81 studies published until June 2021 was 18.1\% (95\% CI: [15.4\%--21.3\%]) \cite{madewell2020household,madewell2021factors}, and showed a weak dependence on household size.  Additionally, our framework can be easily extended to incorporate vaccination, age groups, and other structural levels in contact patterns (e.g., workplaces), which may be relevant for secondary transmission \cite{madewell2022household}.

Our approach to estimating the country-specific COVID-19 prevalence used an estimate of COVID-19-related deaths, as seroprevalence studies at the population scale and across all regions we analyzed were unavailable, e.g., in \cite{arora2021serotracker}.  
There are several studies dedicated to the estimation of worldwide COVID-19-related deaths, e.g., \cite{wang2022estimating,msemburi2023estimates}; however, there are strong discrepancies from what local teams have reported (see, e.g., \cite{bager2023conflicting,moeti2023conflicting,oNeill2023conflicting,Scholey2023conflicting}). We conducted a comprehensive analysis and parameter scanning in the Supplementary Materials to confirm that our results remain valid within the reported parameter and variable ranges. We also tested and verified the reliability of our results using alternative methodologies and provided original data to support our parameter choices.

% Outlook
To sum up, household structure, a cornerstone of cultural diversity worldwide, is a resilient entity that endures despite the influence of contemporary economic and political factors. Understanding the structural challenges that countries and regions face due to their household structures and their potential to accelerate epidemics is an urgent need in our efforts to enhance preparedness and effectively respond to future threats to global public health.

\section*{Methods}

\subsection*{The relation between prevalence, household structure, and out-household reproduction number}
\label{sec:methods-relations}

We consider a network SIR (susceptible-infectious-recovered) epidemic framework with households and possible time-varying out-household contact structures. Each infected individual in the network has an unknown out-household reproduction number $\rho_{i}$ at the time the individual became infected, where $i$ indexes the cumulative set of infected individuals at time t, $I$ (for ease of notation, we will ignore the explicit time dependence). The term \textit{infected} involves the currently infectious and the recovered (assumed to be immune). We also assume a household-size-independent in-household secondary attack rate $a_{H}$, although the theory remains valid for household-size-dependent attack rates. 

Let the population be denoted by $n$ ($n\gg1$), $H_{k} $ the number of households of size $k$, and $H=\sum H_{K}$ the total number of households. The household size distribution $\eta$ is thus defined as $\eta_{k}=\frac{H_{k}}{H}$. Let further $E\left(\eta\right)=\sum\limits_{k} k\eta _{k}$ be the average household size. It follows that $\sum\limits_{k}kH_{k}=n=H\cdot E$ and hence $H=\frac{n}{E}$. First, we compute the probability that an individual had an infectious out-household contact, $p_{\rm out}$. Note that having an infectious out-household contact does not mean the individual was infected by that contact, as they may already have been infected earlier by an infectious in-household contact. Let $\alpha$ denote the prevalence in the population, which, under the assumption of perfect immunity after infection, can be expressed as $\alpha \stackrel{\text{def}}{=} \frac{I}{n}$. For $n\rightarrow \infty $ we obtain the probability for an infectious out-household contact $p_{\rm out}$ as

\begin{align}
p_{\rm out} &=1-\prod\limits_{i\in I}\left( 1-\frac{\rho_{i}}{n}\right)  \\
&\simeq 1-\exp \left( -\frac{1}{n}\sum \rho_{i}\right)  \\
&=1-\exp \left( -\frac{\alpha }{I}\sum \rho_{i}\right)  \\
&=1-\exp \left( -\alpha \cdot R_{\rm out}\right)\,, \label{eq:sup-relation-rout-po}
\end{align}

where $\avOutHouseRepNumI = \frac{\sum_{i\in I} \rho_i}{I}$ is the average of the individual $\rho_i$.

In the following, we show how the out-household infection probability $p_{\rm out}$ and the household secondary attack rate $a_H$ (more generally, the assumptions on the in-household transmission probabilities) determine the expected number of infected household members.  Let $\mu _{k}=\mu _{k}\left( p_{\rm out}\right)$ represent the expected number of infected individuals in a household of size $k$.  Furthermore, $\eta_k$ and $\mu_{k}$ are directly related to the prevalence $\alpha$, enabling the computation of the out-household reproduction number $\avOutHouseRepNumI$, given $a_H$. We have, 

\begin{align}
\sum\limits_{k\geq 1}H_{k}\mu _{k} &=\alpha \cdot n \\
&\text{and hence} \\
\sum\limits_{k\geq 1}\eta _{k}\mu _{k} &=E\cdot \alpha \,.
\end{align}%

For $\mu _{k}\left( p_{\rm out}\right) $ we have 

\begin{equation}
\mu _{k}\left( p_{\rm out}\right) =\sum\limits_{l=1}^{k}d_{l,k}\binom{k}{l}%
p_{\rm out}^{l}\left( 1-p_{\rm out}\right) ^{k-l}\,,
\end{equation}

where $d_{l,k}$ is the expected number of infected individuals in a household of size $k$ if $l\geq 1$ household members were exposed to infection from outside the household. In general, $d_{l,k}$ can be expressed via recursions. In the following, we illustrate a possible way to compute $d_{l,k}$ when the in-household infection process is described via an Erdős-Rényi random graph model with edges representing infections with edge probability $p_{k}=p\left( k\right)$ (which might be different for different household sizes $k$). To compute $d_{l,k}$, we have to estimate the probability that exactly $m$ nodes are connected to at least one of the fixed $l$ externally exposed household members. Let $P_{l}\left( m,k\right) $ be this probability. Let further $Q_{l}\left( k,p_{k}\right) $ be the probability that in a household of size $k$ with $l$ external exposures, all household members will become infected. As a natural consequence, $Q_l(l,p_k)=1$ always holds. We have 

\begin{equation}
Q_{l}\left( k,p_{k}\right) =1-\sum\limits_{m=l}^{k-1}\binom{k-l}{m-l}%
Q_{l}\left( m,p_{k}\right) \left( 1-p_{k}\right) ^{m\left( k-m\right) } \label{eq:Qlk-recursion}
\end{equation}%
and 
\begin{equation}
P_{l}\left( m,k\right) =\binom{k-l}{m-l}Q_{l}\left( m,p_{k}\right) \left(
1-p_{k}\right) ^{m\left( k-m\right) }\,.
\end{equation}

Finally, we get 
\begin{equation}
d_{l,k}=\sum\limits_{m=l}^{k}mP_{l}\left( m,k\right)\,. 
\end{equation}

To start the recursion, one starts with $Q_l(l,p_k)=1$ and proceeds by iteratively computing $Q_l(l+i,p_k)$ for $i=1,2,...$ (up to households of size 6, as the Eurostat household size distributions are truncated at that value \cite{EurostatHHdist}) using Eq.~\ref{eq:Qlk-recursion}. With the $Q_l(l,p_k)$ one proceeds to compute the $P_{l}\left( m,k\right)$ and $d_{l,k}$. With the $d_{l,k}$ one can compute the $\mu_k(p_{\rm out})$, via which one solves $\sum\limits_{k\geq 1}\eta _{k}\mu_{k}=E\cdot \alpha $ for $p_{\rm out}$. Via Eq.~\ref{eq:sup-relation-rout-po}, this offers a relation between the prevalence $\alpha$, the household size distribution $\eta$, and the out-household reproduction number $\avOutHouseRepNumI$. 

Lastly, to estimate the values of $p_{k}$ for different household sizes, we choose a fixed in-household secondary attack rate $a_{H}=0.2$ (assumed to be independent of the household size $k$), defined as the expected number of secondary cases generated from a first, single infection of the household from outside. Hence, we solve the above recursion equations for $d_{l,k}$ such that $a_{H}=\left( d_{1,k}-1\right)
/\left( k-1\right) $ is equal to a constant $a_H$. Once the $d_{l,k}$ have been computed as a function of the $p_{k}$, one can solve the fixed point equation $\sum\limits_{k\geq 1}\eta _{k}\mu_{k}=E\cdot \alpha $, which has a nontrivial solution for $p_{\rm out}$ and hence for $\avOutHouseRepNumI$ (Eq.~\ref{eq:sup-relation-rout-po}) when the household reproduction number (defined in the next section, Eq.~\ref{eq:crit_suppl}) is $\RH>1$.

\begin{table}[htp]\caption{\textbf{Model variables.}}
\label{tab:variables}
\centering
\begin{tabular}{lp{6cm}p{7cm}}\toprule
\makecell[l]{Variable} & Name & Formula  \\\midrule
$\rho_i$ & Individual out-household reproduction number  & -- \\
$\bar R_{\rm out}$ & Average out-household reproduction number  & $\bar R_{\rm out} = \frac{1}{I}\sum \rho_i$ \\
$\momratio$& Effective household size & $\mathrm{E}\left(\eta\right)+\frac{\mathrm{Var}\left(\eta\right)}{\mathrm{E}\left(\eta\right)}$\\
$\alpha$ & Prevalence & -- \\
$\Boost$ & Household boost factor & $\mathcal{B} = 1+a_H \left(\momRatio-1 \right)$ \\
$\mathcal{R}_H$ & Household reproduction number & $\mathcal{R}_H = R_{\text{out}, t} \mathcal{B}$ \\
\bottomrule
\end{tabular}%
\end{table}

\begin{table}[htp]\caption{\textbf{Model parameters.}}
\label{tab:parameters}
\centering
\begin{tabular}{llcl}\toprule
\makecell[l]{Parameter} & Name & Value (default, range)  & Source \\\midrule
$a_H$ & In-household secondary attack rate & 0.2 [0.18 -- 0.22] & \cite{madewell2022household,madewell2020household}\\
$\gamma$ & Fraction of excess deaths attributable to COVID-19 & 0.85 [0.70 -- 1.00] & Estimated \\
$\eta$& Household size distribution & country specific & \cite{EurostatHHdist}\\
\bottomrule
\end{tabular}%
\end{table}
\clearpage
\subsection*{The household reproduction number $\mathcal{R}_H$ } 

Here, we describe the initial infection process between households and derive the equation for the household reproduction number. The phase transition for the emergence of a giant component in random graphs with household structures and its relation to epidemics has already been studied in great detail in the works of F. Ball (see, e.g., \cite{PELLIS201285}). Here, we give a short derivation of the equation for the household reproduction number based on the theory of inhomogeneous random graphs.  We consider a directed random graph model where edges stand for potential infectious contacts, where the probability that an infected individual $i$ will infect an individual $j$ is given by 

\begin{equation}
\Pr \left( i\leadsto j\right) =\frac{\rho_{i}}{n}\,.
\end{equation}

We assume that edges are independent, the values $\rho_i$ are independent of the household size, and the $\rho_i$ distribution has an asymptotic limiting distribution for large \textit{n} with finite expectation \avOutHouseRepNumRH. The network defined by this edge formation rule would be a directed Bollobas-Janson-Riordan graph \cite{https://doi.org/10.1002/rsa.20168}, also known as an inhomogeneous random graph, if some mild conditions on the $\rho_i$ distribution were imposed, e.g., boundedness or a finite first moment. Additionally, the population is partitioned into households. Hence, the out-household network structure induces a network structure among the households. We can describe the household connectivity structure using a kernel that specifies the probability that an infected household of size $x$ is connected to a household of size $y$ (note that all edges represent infectious events). It is given by the household contact kernel $\varkappa_{H}:$

\begin{eqnarray}
\Pr \left( x\leadsto y\right) &=&\frac{\varkappa _{H}\left( x,y\right) }{H}
\\
\varkappa _{H}\left( x,y\right) &=&\frac{\avOutHouseRepNumRH}{E\left(\eta\right)}\left( 1+a_{h}\left(
x-1\right) \right) \cdot y,
\end{eqnarray}

The connectivity properties of the network of households determine the epidemic threshold. The threshold criterion can be formulated as the condition that the household-reproduction number $\RH$ (defined as the average number of secondary households infected by an infected household) equals one. Applying the general theory for directed Bollobas-Janson-Riordan graphs \cite{https://doi.org/10.1002/rsa.20892} (see also the appendix in \cite{doi:10.1137/23M1556861}), we obtain the household reproduction number

\begin{equation}\label{eq:crit_suppl}
\RH=\avOutHouseRepNumRH \left(1+a_{H}\left( \momratio-1\right) \right)\,,
\end{equation}%

where the effective household size $\eta^*$ emerges as the quotient between the second and first statistical moments of $\hhDist$. 

We close this section by providing a relation between the out-household reproduction number $\avOutHouseRepNumRH$ at a given time and the average out-household reproduction number \avOutHouseRepNumI, averaged over all infected individuals in an outbreak. Recall that $\avOutHouseRepNumRH$ is the mean number of secondary cases generated by newly infected individuals at time $t$ (here, we consider $t$ to be a discrete-time unit, e.g., a day or a week, short enough that changes of $\avOutHouseRepNumRH$ within such a time unit can be neglected). Let furthermore $I_t$ be the newly infected individuals at time $t$. We then have the relation

\begin{equation}
\avOutHouseRepNumI=\sum\limits_{t=0}^{T}\avOutHouseRepNumRH \cdot  I_t / \sum\limits_{t=0}^{T} I_t\,.
\end{equation}

\subsection*{Partial correlation analysis}

The estimated prevalence $\alpha$ correlates with socioeconomic indicators, such as the Human Development Index (HDI) and its components. However, both quantities are (directly or indirectly) affected by other variables, such as the effective household size $\momRatio$ or the out-household reproduction number $\avOutHouseRepNumI$. We aim to quantify how much of the correlation between prevalence and the HDI remains after excluding the contributions of those variables. The semi-partial correlation quantifies how much of the correlation remains between two variables (HDI and prevalence $\alpha$) when the influence of a third variable is removed from only one of them ($\momRatio$ or \avOutHouseRepNumI from COVID-19 prevalence) \cite{baba2004partial}. In contrast, the partial correlation quantifies how much of the correlation remains after controlling for the variable that influences both variables. The following paragraph outlines how to compute both the partial and semi-partial correlations. 

Generally, if the relationships among all variables are linear, the partial correlation can be computed as follows. Let $X=(x_1,...,x_N)$ and $Y=(y_1,...,y_N)$ be the correlating variables, and $Z=(z_1,...,z_N)$ be the confounding variable that influences both $X$ and $Y$. First, one conducts two separate linear regressions with the confounding variable, namely one between $X$ and $Z$ and the other between $Y$ and $Z$. Let $e_X$ and $e_Y$ be the residuals of the two linear regressions. The partial correlation $r_{XY,Z}$ is then given by the Pearson correlation coefficient between the residuals $e_X$ and $e_Y$. As the residuals represent the relationship not accounted for by the confounding variable, this method excludes its contribution. In case the confounding variable only influences $X$, the semi-partial correlation can be computed as the Pearson correlation between the residuals $e_X$ with the variable $Y$ itself, i.e., without performing the linear regression between $Y$ and $Z$.

In our case, the semi-partial correlation is computed as the Pearson correlation between the residuals $e_\alpha$ and the HDI. However, because our framework provides a nonlinear theoretical relationship between the prevalence $\alpha$ and the confounding variables, the residuals $e_\alpha$ are not computed via linear regression but rather using the theoretical prevalences as described in the main text. When the confounding variable is $\momRatio$, the theoretical prevalences are computed assuming a common \avOutHouseRepNumI across countries. Vice versa, when the confounding variable is \avOutHouseRepNumI, theoretical prevalences are computed under the assumption of a constant household distribution across countries. The residuals are then correlated with the HDI across all samples for each IFR vector, yielding a distribution of semi-partial correlation coefficients.

\subsection*{Data sources}

Prevalence estimates are based on annual population size, weekly deaths from all causes, and reported COVID-19 deaths (for details, see SI, Section~\ref{sec:seup-extended-deaths}). The first two datasets, covering 2015–2021, come from Eurostat. However, they are incomplete for some countries; for instance, overall death data are missing for Ireland, Luxembourg, and North Macedonia. Without complete all-cause mortality data, excess deaths cannot be estimated. Nevertheless, prevalence for these countries can still be inferred from officially reported COVID-19 deaths. To this end, we used a dataset of daily reported COVID-19 deaths from Johns Hopkins University, which covers all European countries, and aggregated it to weekly counts.

\clearpage
%TC:ignore
\section*{Acknowledgements} 

We thank the Priesemann group for the exciting discussions and their valuable input. Authors with affiliation "1"  received support from the Max-Planck-Society. SC, PD, VP, and TK were funded by the German Federal Ministry for Education and Research for the RESPINOW project (031L0298G), and SC, VP, and JW by the infoXpand project (031L0300A). VP was supported by the Deutsche Forschungsgemeinschaft (DFG, German Research Foundation) under Germany’s Excellence Strategy - EXC 2067/1-390729940. VP also acknowledges the support of the Ministry of Science and Culture of Lower Saxony through funds from the program zukunft.niedersachsen of the Volkswagen Foundation for the 'CAIMed – Lower Saxony Center for Artificial Intelligence and Causal Methods in Medicine' project (grant no. ZN4257), ``Niedersächsisches Vorab'', and the Niedersachsen-Profil-Professur. TK acknowledges support from the Government of Saxony through the SaxoCOV grant. Grammarly Pro AI has been used for grammar checks. 

\section*{Author Contributions}
% Using the CReDiT statement.
%\scc{see details here: \url{https://www.elsevier.com/researcher/author/policies-and-guidelines/credit-author-statement}}

Conceptualization: JPB, TK, VP \\
Data curation: FR, JW, MF, PD\\
Formal analysis: JPB, JW, MF, PD, TK, VB\\
Funding acquisition: SC, TK, VP\\
Investigation: BP, FR, JPB, JW, MB, MF, PD, SC, VP\\   
Methodology: FR, JPB, MB, PD, TK \\
Project administration: SC, TK, VP\\
Resources: TK, VP\\
Software: JPB, JW, MF, PD, JPB  \\
Supervision: SC, TK, VP\\	
Validation: all\\
Visualization: JW, MF, PD, SC \\	
Writing - Original Draft: BP, FR, JPB, JW, MF, PD, SC, TK, VP\\	
Writing - Review \& Editing: all \\

%\renewcommand{\refname}{References (1--10 for the main text, 1--39 for the arXiv version)}
%\bibliographystyle{unsrt}
%\bibliography{references}

%%% Supplementary Materials!
\newpage

%%% Preamble
\renewcommand{\thefigure}{S\arabic{figure}}
\renewcommand{\figurename}{Supplementary~Figure}
\setcounter{figure}{0}
\renewcommand{\thetable}{S\arabic{table}}

\setcounter{table}{0}
\renewcommand{\theequation}{\arabic{equation}}
\setcounter{equation}{0}
\renewcommand{\thesection}{S\arabic{section}}
\setcounter{section}{0}

\input{supplementary_material}

%TC:endignore
\end{document}

%% file: supplementary_material.tex
\section*{Supplementary Material}

The Supplementary Material provides an in-depth overview of our research methodology, including detailed analysis methods and data sources. It includes a sensitivity analysis that assesses the robustness of our results to modeling assumptions, data source uncertainties, and key parameters. Additionally, an expanded sociological analysis of our findings is presented, along with supplementary figures and tables that display the data referenced in the manuscript and define relevant variables and parameters.

\tableofcontents

\clearpage
\section{Extended methods}
\label{sec:supplementary-S1-1-model}
\subsection{Prevalence estimation}

Estimating the prevalence $\alpha$ is a challenging task. For example, reporting practices and capabilities vary widely across countries, leading the number of officially reported COVID-19 cases to deviate substantially from the true number of cases. A more reliable method for obtaining a prevalence estimate involves using death reports and country-specific infection fatality rates (CIFRs). Given a death and a CIFR estimate, the prevalence $\alpha$ can be estimated by 

\begin{equation}\label{eq:pravalence}
    {\alpha} = \frac{D}{\text{CIFR}},
\end{equation}

where $D$ is the number of deaths. Our approach to estimating the true prevalence consists of two steps. First, we estimate the CIFRs based on \cite{brazeau2020report,brazeau2022estimating}. Second, we estimate the true number of COVID-19 deaths using observed surplus and reported COVID-19-related deaths. We cover the period from the start of the pandemic until June 13th, 2021. We chose this end date for two reasons. First, it is early enough that mass vaccinations have not yet had a significant effect on reducing COVID-19-related deaths. Second, it represents a date at which infection numbers were low across all European countries \cite{owidcoronavirus}.

\subsubsection{Estimation of country-specific infection fatality rates (CIFRs)}

Age is one of the dominant factors for COVID-19 severity and fatality risk \cite{odriscoll_age-specific_2021,Levin2020,brazeau2022estimating}. The marked exponential dependency of COVID-19 infection fatality rate (IFR) on age has some noise in the extremes of the very young and very old. We used the age-dependent IFR and confidence intervals reported in \cite{brazeau2022estimating} to reconstruct the uncertainty curve, i.e., estimate the parameters of the uncertainty log-normal distribution around each age point. 

% new version equal to code.
Let $s=1,\ldots,N$ be the age groups used for reporting the age-dependent $\text{IFR}^s$ and $A^s$ the collection of discrete ages encompassed in age group $s$. For example, if age group $k$ represents the age range $[a^k_1,a^k_2]$, then $A^k = \left\{a^k_1,\,a^k_1+1,\,...,\,a^k_2-1,\,a^k_2\right\}$. Let further
\begin{equation}
    \xi^* = \underset{\xi}{\operatorname{argmin}} \sum\limits_{s = 1}^{N} \left(\dfrac{\text{IFR}^s\sum\limits_{i\in A^s} p_i - \sum\limits_{i\in A^s} p_i f_{\text{IFR}}(i;\xi) }{\sum\limits_{i\in A^s} p_i}\right)^2
\end{equation}

% new version equal to code.
where $\xi = [\begin{array}{cccc} \alpha & \beta & \gamma & \delta \end{array}]$ and
\begin{align}
    f_{IFR}(i;\xi) =& e^{\alpha h(i;\xi) + \beta}\label{eq:ifr}\\
    h(i;\xi) =& \frac{i}{2} -  \frac{1}{2\gamma}\log{\left(\frac{\cosh{[\gamma(i - \delta)]}}{\cosh{[\gamma(- \delta)]}}\right)}\,.\label{eq:h}
\end{align}

These functions were selected to capture the exponential dependency of IFR on age, with saturation for high ages (above 90 years). In this context, Eq.~\ref{eq:h} represents the saturation component of the exponent in Eq.~\ref{eq:ifr}. For a given sample mortality curve parameter vector ($\xi$) (i.e., the result of reconstructing the mortality curve ($f_{\text{IFR}}$) from the uncertainty distributions at each point), we can estimate the country-specific IFR (scalar) as a weighted average using the above formula as follows

\begin{equation}
    \text{CIFR} = \sum\limits_{i=0}^{M} p_i f_{\text{IFR}}(i;\xi),
\end{equation}

where $p_i$ is a fraction of the population at age $i$ in a given country and $M$ is the maximum age in the population. Note that the mortality curve is the same for all countries (for the same params vector $\xi$) and that the differences in CIFRs between countries are due to the population's age profile (demographics).

\begin{figure}
    \centering
    \includegraphics[width=12cm]{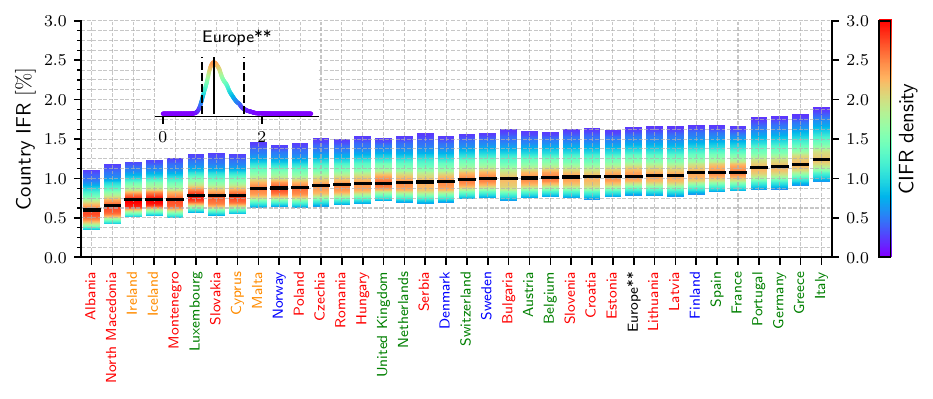}
    \caption{\textbf{Country infection fatality rate (CIFR) estimation for different countries.} We compute country-specific IFRs using age-resolved estimates of the IFR  \cite{brazeau2020report, brazeau2022estimating} and country demographics. Given that age-resolved IFRs are reported as a distribution, we draw 50,000 samples and fit an exponential function to capture the correlation between measurements. We use the fitted profile as an estimated CIFR and weight it by country demographics to obtain a country-specific IFR (color scheme). The vertical bars extend until the 10th and 90th percentiles, and the black bar indicates the median. Europe marked with a double asterisk (**) includes only countries for which we have full Eurostat data.}
\end{figure}

\subsubsection{Estimation of $\gamma$-extended COVID-19 deaths}
\label{sec:seup-extended-deaths}

\begin{figure}[htbp]
    \centering
    \includegraphics[width=12cm]{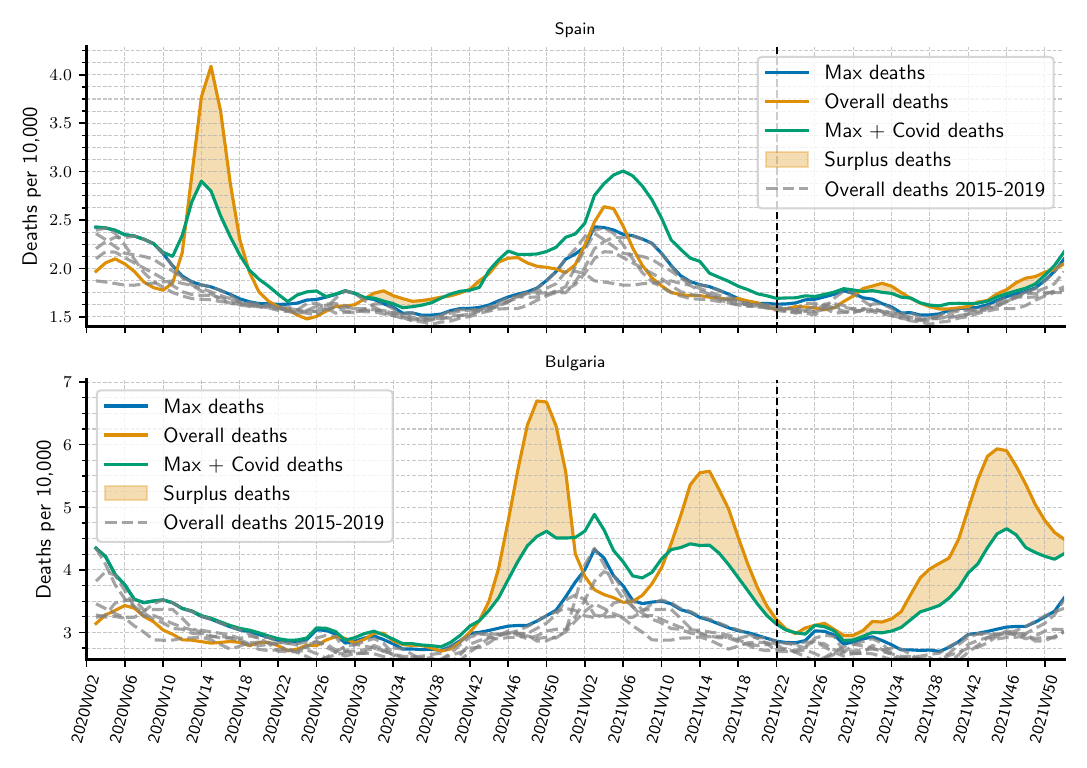}
    \caption{\textbf{Example estimation of total COVID-19 deaths for Bulgaria and Spain.} The blue curve shows the maximal weekly deaths over 2015-2019. It is added to the officially reported COVID-19 deaths (green curve). If the overall deaths in any given week during the pandemic exceed this green curve (orange shaded area), we attribute a fraction $\gamma$ of these additional deaths as being COVID-19 related. This way, we obtain the $\gamma$-extended COVID-19 death estimate. The dashed line represents the end date of our $\gamma$-extended deaths and prevalence estimations.}
    \label{fig:S2-surplusdeaths_bulgaria-spain}
\end{figure}

Determining the true number of COVID-19-related deaths $D$ is challenging, e.g., because many COVID-19 reported deaths were not officially counted as such. To find an accurate estimate, we thus add the officially reported COVID-19 deaths to the maximum number of weekly deaths over previous years --- if we observed even more deaths than that, we attribute a fraction $\gamma$ of those to being COVID-19 related (Fig.~\ref{fig:S2-surplusdeaths_bulgaria-spain}). We call this estimate the $\gamma$-extended number of COVID-19 deaths. Similar adjustments are sometimes also known as \textit{excess deaths} \cite{owidcoronavirus}.

In particular, we use the following procedure. 

\begin{enumerate}
    \item We normalize deaths $D_a(T)$ in the year $J$ using population size $P_a(J)$
        \begin{equation*}
            d_{a}\left( T\right) := D_{a}\left(T\right) / P_{a}\left( J\right).
        \end{equation*}
    \item We find the maximum number of deaths in each week $w$ over previous years
        \begin{equation*}
            d_{a}^{\rm max}\left( w\right) := \underset{J \in [2015,2019]}{\max} \left\{d_{a}\left( T\right)\right\}.
        \end{equation*}
    \item Using a three-week moving average, we smoothen the maximum deaths 
        \begin{equation*}
            d_{a}^{\ast }\left( w\right) :=\frac{1}{3}\left( d_{a}^{\rm max}\left( w-1\right) +d_{a}^{\rm max}\left( w\right) +d_{a}^{\rm max}\left(w+1\right) \right). 
        \end{equation*}
    \item We calculate surplus deaths $S_{a}\left( T \right)$ as
        \begin{align*}
            S_{a}\left( T\right) :=& \max\left\{ 0,P_{a}\left( J\right) \cdot \max \left\{ d_{a}\left( T\right) -d_{a}^{\ast }\left( w\right), 0 \right\} -C_{a}\left( T \right) \right\}
        \end{align*}
    \item For a specific value of $\gamma$, we calculate $\gamma$-extended COVID-19 deaths $E_{a}\left( T,\gamma \right)$:
                \begin{equation*}
            E_{a}\left( T,\gamma \right) := C_{a}\left( T \right) +\gamma \cdot S_{a}\left( T\right) .
        \end{equation*}
\end{enumerate}

\begin{figure}[htbp]
    \centering
    \includegraphics[width=12cm]{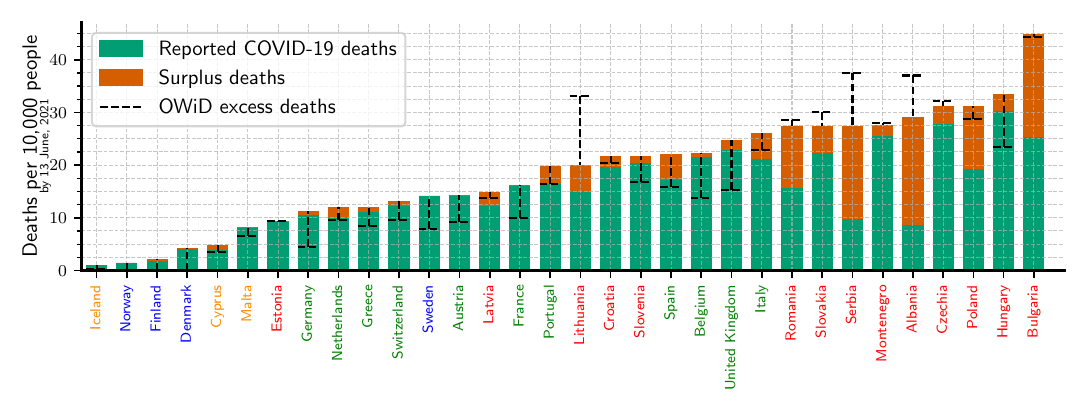}
    \caption{The officially reported number of COVID-19 deaths, surplus deaths, and $\gamma$-Extended deaths, estimated as described in the main text. Cumulative officially reported COVID-19 deaths are represented by green bars, and cumulative excess deaths are represented by orange bars. The sum of these bars corresponds to the number of $\gamma$-extended COVID-19 deaths where $\gamma$ equals 1. For countries marked with an asterisk (*), the excess of deaths is 0 due to incomplete Eurostat data. Europe marked with a double asterisk (**) uses aggregated data from countries without an asterisk (all data were available).}
    \label{fig:S2-excessdeaths_barchart}
\end{figure}

This procedure is visually represented in Fig.~\ref{fig:S2-surplusdeaths_bulgaria-spain}. Compared to the officially reported COVID-19 deaths, this method substantially increases the death estimate in some countries (Fig.~\ref{fig:S2-excessdeaths_barchart}).

\begin{figure}[htbp]
    \centering
    \includegraphics[width=12cm]{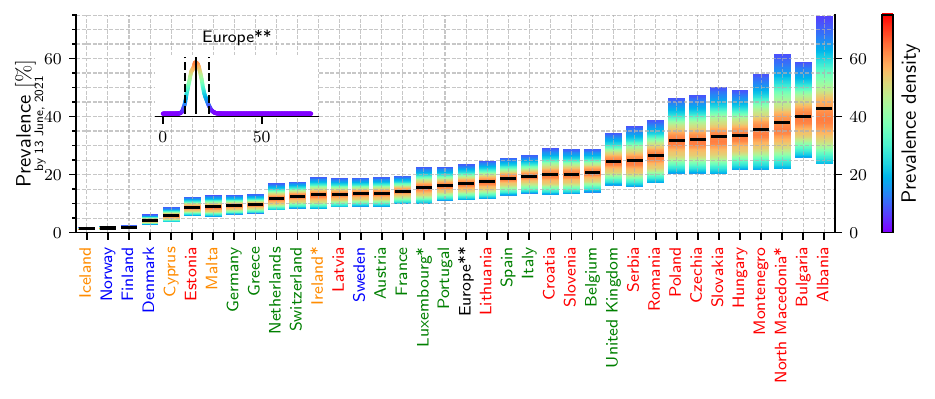}
    \caption{\textbf{Prevalence estimation for different countries.} We obtain a distribution over the prevalence by Using the estimated COVID-19 deaths and the country infection fatality rate (CIFR) estimates. Vertical bars extend to the 10th and 90th percentiles, and the black bar indicates the median. For countries marked with an asterisk (*), the excess of deaths is 0 due to incomplete Eurostat data. Europe, marked with a double asterisk (**), uses aggregated data from countries without an asterisk (all data were available).}
    \label{fig:S2-distribution-prevalence}
\end{figure}

Combined with the IFR estimates, we thus obtain a distribution over prevalences (Fig.~\ref{fig:S2-distribution-prevalence}).

\subsubsection{Estimation of $\gamma$}

One of the central parameters in our analysis is $\gamma$, i.e., the fraction of the surplus deaths attributable to COVID-19. However, this parameter is hard to estimate. Analyses of the availability and delivery of healthcare services during the pandemic support the hypothesis that excess mortality during the pandemic was mainly attributed to COVID-19-related infections. 

Overall, excess mortality coincided with successive waves of infection, during which healthcare systems operated under unusual conditions. Mortality during these periods may have been related to direct SARS-CoV-2 infections or indirect effects related to healthcare access and delivery. To assess this, we analyzed data from the Polish National Health Fund on urgent medical interventions, including procedures such as treatment of stroke, heart attack, and aneurysm perforation. These services, if left untreated, usually lead to mortality within timescales comparable to waves of infection.

We use the following procedure to estimate the fraction $\gamma$ of additional deaths that we attribute to being COVID-19 related.
\begin{enumerate}
    \item We find the mean number of emergency interventions in each week $w$ over previous years and smooth them
    \begin{equation*}
        \text{BL}(w) = \frac{1}{5}\sum_{j = 2015}^{2019}\frac{1}{3}(\text{EI}(w - 1, j) + \text{EI}(w, j) + \text{EI}(w + 1, j))
    \end{equation*}
    \item We calculate the lacking procedures during the weeks of the pandemic 
    \begin{equation*}
        \text{LP}(w, J) = \max\{0, \text{BL}(w) - \text{EI}(w, J)\}
    \end{equation*}
    \item We estimate the parameter $\gamma$ using the surplus deaths $S(w)$
    \begin{equation*}
        \gamma(w, J) = \frac{\sum_{j=2020}^{J}\sum_{w}^{N_j}\max\{0, S(w) - \text{LP}(w, j)\}}{\sum_{j=2020}^{J}\sum_{w}^{N_j}S(w)}
    \end{equation*}
    where $N_j=w$ if $j=J$ and $N_j=52$ if $j<J$\,.
\end{enumerate}

\begin{figure}[htbp]
    \centering
    \includegraphics[width=12cm]{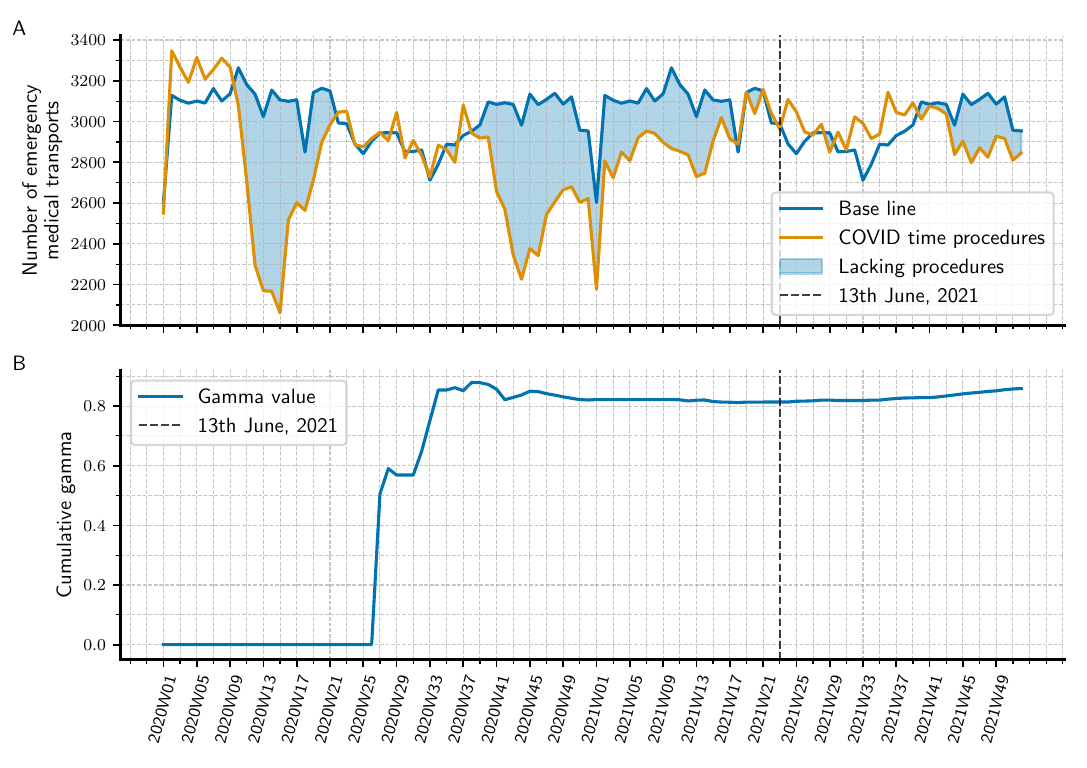}
    \caption{\textbf{A.} Number of emergency procedures due to all causes except COVID-19 in Poland, before and during the pandemic. \textbf{B.} Cumulative deviation from historical trends, which is a proxy for $\gamma$.}
    \label{fig:gamma_estimates}
\end{figure}

Fig.~\ref{fig:gamma_estimates} shows a visual representation of this procedure. We estimate $\gamma$ to lie in the range of $\gamma \approx 0.85$, which we used in the analysis in the main text. 

\subsubsection{Comparison of excess mortality estimates and $\gamma$-extended COVID-19 deaths}

Estimates of excess mortality from Our World in Data are calculated using a linear model \cite{karlinsky2021tracking}. The model is based on linear regression fitted to historical weekly mortality data from 2015-2019, which provides a baseline for expected deaths in subsequent years. Excess mortality is the cumulative difference between observed deaths in a given week and the model's predicted baseline for that week. This approach allows negative excess mortality in some weeks, which often yields lower excess mortality estimates than our method. In addition, the model does not include officially reported COVID-19 deaths in its estimates, which can result in even lower COVID-19 excess mortality estimates, particularly at the beginning of the pandemic.

\section{Social determinants of household size}
The COVID-19 pandemic has highlighted the social context's essential role in shaping responses to health crises, particularly that of households. As fundamental social units, households influence not only the transmission of infections but also the beliefs, behaviors, and coping strategies that guide individual responses to health emergencies. This study positions the household as the primary unit of analysis, recognizing that its diversity in size and structure is critical for understanding the dynamics of infection spread.
 
As illustrated in Fig.~\ref{fig:critical-outhousehold}, the effective household size \momRatio (i.e., mean household size plus a --positive-- variance-dependent correction factor) correlates with the out-household reproduction number \avOutHouseRepNumI values. To understand how household size affects \avOutHouseRepNumI and may shape pandemic-response patterns, we must view household structures as outcomes of long-run economic and cultural processes. A global trend toward smaller households has been evident for centuries, particularly in developed countries \cite{Bradbury2014}. This trend is particularly pronounced in Europe, though significant regional variation in household composition remains \cite{Esteve2024,IUSSP2023} (Esteve et al., 2023; N-IUSSP, 2023). Household structure across Europe can be categorized into four broad regions: Western, Eastern, Southern, and Nordic Europe. This distinction, first conceptualized by Frédéric Le Play in the 1870s and later popularized by John Hajnal in the 1960s, reflects enduring patterns of family composition \cite{LePlay1877,Hajnal1965}. Studies support this categorization, with Nordic countries exhibiting a high prevalence of single-person households, Western Europe dominated by nuclear families, Southern Europe characterized by larger family units, and Eastern Europe showing extended family systems \cite{Laslett1983,Hajnal1982,Kalmijn2007,Saraceno2008,Reher1998,Iacovou2004,szoltysek2012spatial,Iacovou2011}.

There is a strong correlation between household size and the HDI (Fig.~4A), particularly with the economic component (GNI). These West/North-East/South differences in household size can be analyzed using modernization theory. This theory posits that as societies undergo economic development, they experience structural and cultural changes that transform social institutions, including family structures, which directly affect household size (urbanization, industrialization, income growth, mobility) \cite{Giddens1990,vanDenBan1967}. Economic growth, a core element of modernization, promotes the emergence of more independent family units, leading to smaller households. As traditional values give way to individualistic cultural frameworks, societies shift from multigenerational households to nuclear families or single-person households \cite{Iacovou2011,Esteve2024}. While HDI does not directly measure cultural patterns, it indirectly reflects cultural transformations driven by shifts in the economy and society. For example, the expansion of women's education increases their income potential, promotes financial independence, shapes career aspirations, often leads to lower fertility rates, and ultimately  changes the  accepted family model. 

The size and structure of households are shaped by a combination of economic, cultural, and institutional factors. These same factors not only determine household size but also affect how societies respond to pandemics. For instance, factors like housing conditions, employment structures, and welfare support systems influence both household  size and the ability to implement health measures such as isolation and social distancing. By analyzing the determinants of household size, we also gain insight into the broader mechanisms that shape pandemic risk and public health outcomes. These factors will be briefly discussed.

\subsection{Economic factors}

Economic conditions, particularly income levels and housing costs, shape household size across Europe.  Wealthier regions like Western and Northern Europe exhibit smaller households, often nuclear families or single-person units, driven by financial independence and strong social safety nets. Countries such as Switzerland and Luxembourg, which report the highest GNI and disposable incomes, illustrate this trend. Even in Northern Europe, where welfare systems reduce disposable income (e.g., Norway: \$32,514; Denmark: \$25,859), household sizes remain small due to comprehensive welfare systems. In contrast, Southern and Eastern Europe face greater economic challenges, such as lower GNI and higher unemployment, housing cost and poor welfare system leading to larger household sizes. Families in countries like Italy (GNI: \$39,294) and Greece (GNI: \$21,391) often rely on shared resources and multigenerational living due to limited financial independence. This pattern is most prominent in Eastern Europe, where lower disposable incomes, as seen in Romania (\$12,635) and Bulgaria (\$12,092), drive families toward shared living arrangements. Data reported in \cite{OECD2020,Eurostat2025}. Better housing conditions reduce overcrowding in wealthier countries like the Netherlands and Belgium. By contrast, overcrowding rates exceeding in Bulgaria, Latvia, Romania, and Poland (over 1/3 people living in an overcrowded household) result in higher household density, driven by inadequate housing and economic insecurity \cite{Eurofound2019,DuboisNivakoski2023,Eurostat2021}, which may elevate infection risk during pandemics.

Modernization theory explains how differences in economic development between Western/Northern and Eastern/Southern Europe shape employment structures. Consequently, varying levels of economic growth are expected to result in differences in  occupational structure and work organization, which can play a crucial role in pandemic control, i.e., the share of the labor force  working in the industry, services where there is frequent social contact and professionals  with less social contact  and possible remote work. For instance, higher economic development in Western and Northern Europe facilitates greater use of remote work and flexible work arrangements, reducing physical proximity and the risk of infection. Remote work rates in 2020 were highest in Finland (24.82\%), Luxembourg (23.01\%), and Ireland (21.13\%). In contrast, remote work was less common in Southern and Eastern Europe, with Latvia (2.87\%), Romania (3.18\%), and Croatia (3.48\%) reporting the lowest rates. In regions with lower economic development, like Eastern and Southern Europe, employment is often concentrated in site-specific sectors such as manufacturing, construction, and mining, where on-site presence is essential, increasing exposure to infection risks. Occupations requiring physical presence faced higher risks of infection and suffered more significant economic consequences. In contrast, those able to work from home were better protected both health-wise and economically, underscoring the importance of remote work capabilities in mitigating the effects of public health crises \cite{Beland2023,Fadel2022,Brynjolfsson2020,Schmid2022}.

\subsection{Cultural factors}

Cultural factors reinforce the East-South versus West-North divide in household size. Culture values such as embededdness,  collectivism, more pronounced in the East and South, contribute to larger, multigenerational households, while autonomy \& individualist values in the West and North support smaller, nuclear, or single-person households \cite{inglehart2000,schwartz2014}.

There is a reciprocal relationship between culture and household size: cultural values shape family size, and the model of family size reinforces certain cultural orientations. The size of a country's average family or household strongly correlates with cultural values, where larger households emphasize embeddedness, hierarchy, and mastery. These cultural orientations arise from societal norms that prioritize obedience, conformity, and collective interest to maintain order and effective coordination within large family structures. In turn, these norms foster cultural frameworks that support tradition, authority, and group cohesion, making autonomy and egalitarianism less compatible with such settings \cite{schwartz2006}.

Culture also affects the strength of family ties, which is reflected in  frequency of social contacts. For example, young adults in Southern and Eastern Europe often remain in the parental home for longer periods due to stronger social ties. By contrast, young adults in Western and Northern Europe leave home earlier, influenced by cultural norms promoting early independence and institutional support systems for housing and education.
Cultural patterns significantly influence societal beliefs and behaviors, affecting compliance with restrictions during the COVID-19 pandemic \cite{kaeffer2022culture,ibanez2022role,gokmen2021impact,chen2021culture,chen2023impact}.

 For instance,  Gelfand's research indicates that societies with strict social norms ("tight" cultures) experienced fewer COVID-19 cases and deaths compared to more permissive ("loose") cultures \cite{Gelfand2021}. Studies suggest that low trust in governments correlates with higher infection rates \cite{Bollyky2022}.  In Nordic countries, there is a high level of trust, whereas in central-eastern-European countries, levels of trust in government and among individuals are markedly lower \cite{bohle2022east}.  Therefore, the variation in infection rates can be partly attributed to cultural patterns influencing adherence to restrictions and the prevailing levels of trust within a society. 

\subsection{Institutional factors}

Finally, institutional factors, i.e.,  society's forms of organizations, shape household size and influence pandemic outcomes. The North/West vs. East/South divide is reflected in the prevalence of co-residence between older and younger family members and the availability of formalized care facilities for the elderly. When state-provided care is limited, family-based care increases, leading to higher co-residence rates within households. Welfare states in Scandinavia provide support through childcare, eldercare, and housing assistance, enabling individual independence and smaller household sizes. For instance, in Nordic countries such as Norway (highest), Denmark, Finland, and Sweden, over 12\% of the 80+ population reside in formalized care facilities for older people. This rate rises to 16\% in Luxembourg and Switzerland and 24\% in Belgium. In contrast, Poland exhibits much lower rates, with less than 1\% of the 65+ population living in formalized care facilities, increasing only slightly to 1.6\% for the 80+ population (https://nhwstat.org/). State-funded eldercare systems reduce co-residence rates for older adults. For instance, 5\% of those over 65 in Nordic countries live with younger household members (aged 50 or below), compared to 38\% in Poland—the highest rate in Europe \cite{OECD2020,Eurostat2023}. Weaker welfare systems in Southern and Eastern Europe increase reliance on family support, leading to larger, multigenerational households. While eldercare systems reduce co-residence, they also increase the share of older adults living in formal care institutions. During the COVID-19 pandemic, this arrangement posed infection risks in eldercare facilities, highlighting the dual role of welfare systems in shaping household structure and pandemic outcomes.

In summary, prolonged modernization processes---such as shifts in labor market structures, technological advancements, and changes in income distribution---have significantly influenced family models and household sizes. While numerous studies emphasize the impact of cultural factors, our analysis of correlations and partial correlations reveals a discernible pattern: economic and structural factors exert a more substantial influence on social contact structure. We interpret this as follows: structural factors directly shape individual behaviors by limiting opportunities for social interactions (e.g., household size or remote work), whereas cultural factors, being more nuanced, affect social networks in a complex and indirect manner.  

\clearpage
\section{Robustness check}
\label{sec:supplementary-S3-robustness}

Given the uncertainty around the IFR and the fraction of surplus deaths that are to be attributed to COVID-19 ($\gamma$), we proceed iteratively, sampling from the IFR distribution (multivariable) to generate several IFR vectors, which will yield new estimates for the country IFR. We repeat this procedure \num{50000} times and estimate a distribution for the theoretical prevalence according to equation~\ref{eq:pravalence}. This allows us to quantify not only the out-household reproduction number (and thereby have a fair way to compare pandemic management, ruling out the effect of households), but also the uncertainty around this value. The overall methodology is summarized in Fig.~1C.

\begin{figure}[htbp]
    \centering
    \hspace*{-1cm}
    \includegraphics[width = 180mm]{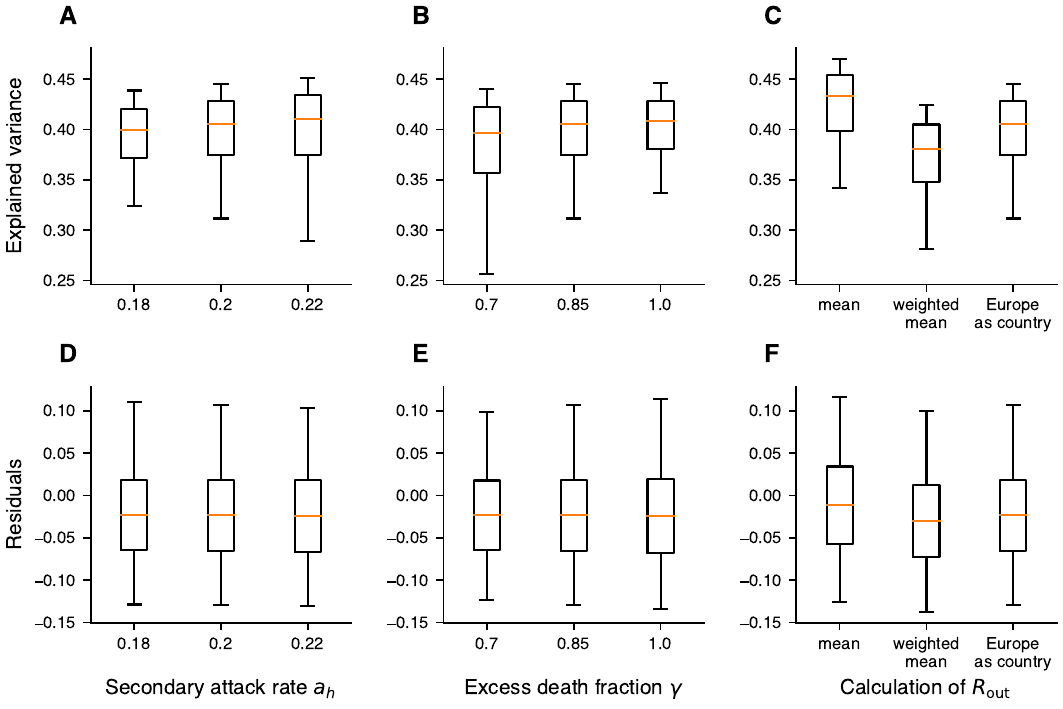}
    \caption{\textbf{Robustness of our results before variations in assumed parameters.} Orange markers denote the median, boxes the interquartile ranges, and whiskers the 10-90 percentiles. }\label{fig:robustness}
\end{figure}

\clearpage

\section{Supplementary Tables}

\begin{table}[htbp]
\caption{\textbf{Data overview.} CIFR denotes the country-infection fatality rate, $D$ the officially reported COVID-19 deaths, $S$ the surplus deaths (Sec.~\ref{sec:seup-extended-deaths}), $\alpha$ the estimated prevalence, $\bar{\eta}$ and \momRatio the mean and effective household size respectively, and $\avOutHouseRepNumI$ the out-household reproduction number. For countries marked with an asterisk (*), prevalence is based on officially reported COVID-19 deaths due to missing data on overall fatalities. Europe (**) uses aggregated data from countries without an asterisk. For distributions, values represent the median and values in brackets the 95\% CI.}
\begin{tabular}{lccccccc}
\toprule
 & CIFR & $D$ & $S$ & $\alpha$ & $E$ & \momRatio & $\avOutHouseRepNumI$ \\
\midrule
\textbf{Albania} & 0.60 [0.38, 1.01] & 8.62 & 20.51 & 42.67 [25.70, 68.81] & 3.63 & 4.23 & 0.87 [0.73, 1.36] \\
\textbf{Austria} & 1.01 [0.79, 1.50] & 14.27 & 0.07 & 13.50 [9.54, 18.13] & 2.20 & 2.93 & 0.79 [0.77, 0.82] \\
\textbf{Belgium} & 1.01 [0.81, 1.50] & 21.57 & 0.64 & 20.64 [14.78, 27.41] & 2.27 & 3.01 & 0.83 [0.78, 0.88] \\
\textbf{Bulgaria} & 1.00 [0.75, 1.53] & 25.42 & 19.41 & 40.02 [27.46, 55.74] & 2.37 & 3.15 & 0.96 [0.84, 1.18] \\
\textbf{Croatia} & 1.02 [0.78, 1.54] & 19.55 & 2.22 & 20.09 [13.95, 27.62] & 2.71 & 3.46 & 0.77 [0.73, 0.83] \\
\textbf{Cyprus} & 0.78 [0.58, 1.22] & 3.92 & 1.01 & 5.84 [3.91, 8.21] & 2.64 & 3.31 & 0.71 [0.70, 0.73] \\
\textbf{Czechia} & 0.92 [0.69, 1.41] & 27.90 & 3.28 & 32.07 [21.79, 44.75] & 2.34 & 2.96 & 0.91 [0.83, 1.05] \\
\textbf{Denmark} & 0.96 [0.73, 1.44] & 4.23 & 0.08 & 4.29 [2.98, 5.86] & 1.98 & 2.64 & 0.77 [0.77, 0.78] \\
\textbf{Estonia} & 1.02 [0.80, 1.52] & 9.36 & 0.08 & 8.74 [6.20, 11.73] & 2.11 & 2.85 & 0.77 [0.76, 0.79] \\
\textbf{Finland} & 1.08 [0.84, 1.58] & 1.74 & 0.37 & 1.83 [1.30, 2.46] & 1.96 & 2.65 & 0.76 [0.76, 0.76] \\
\textbf{France} & 1.08 [0.87, 1.56] & 16.05 & 0.26 & 14.22 [10.40, 18.62] & 2.14 & 2.82 & 0.81 [0.78, 0.84] \\
\textbf{Germany} & 1.15 [0.91, 1.68] & 10.51 & 0.78 & 9.24 [6.64, 12.30] & 2.00 & 2.64 & 0.80 [0.78, 0.82] \\
\textbf{Greece} & 1.18 [0.95, 1.71] & 11.16 & 0.97 & 9.58 [7.00, 12.60] & 2.55 & 3.22 & 0.74 [0.72, 0.76] \\
\textbf{Hungary} & 0.94 [0.71, 1.44] & 30.25 & 3.21 & 33.51 [22.97, 46.33] & 2.30 & 3.02 & 0.92 [0.83, 1.06] \\
\textbf{Iceland} & 0.73 [0.56, 1.15] & 0.97 & 0.00 & 1.25 [0.84, 1.72] & 2.32 & 3.10 & 0.71 [0.71, 0.71] \\
\textbf{Ireland*} & 0.73 [0.54, 1.13] & 9.88 & 0.00 & 13.06 [8.74, 18.21] & 2.62 & 3.38 & 0.74 [0.72, 0.78] \\
\textbf{Italy} & 1.25 [1.00, 1.79] & 21.12 & 5.05 & 19.34 [14.19, 25.37] & 2.30 & 2.95 & 0.82 [0.79, 0.87] \\
\textbf{Latvia} & 1.04 [0.81, 1.55] & 12.33 & 2.51 & 13.20 [9.31, 17.86] & 2.24 & 2.96 & 0.79 [0.76, 0.82] \\
\textbf{Lithuania} & 1.04 [0.82, 1.55] & 14.86 & 5.09 & 17.46 [12.35, 23.55] & 2.16 & 2.83 & 0.82 [0.79, 0.87] \\
\textbf{Luxembourg*} & 0.78 [0.60, 1.22] & 12.86 & 0.00 & 15.55 [10.55, 21.37] & 2.28 & 2.95 & 0.80 [0.77, 0.84] \\
\textbf{Malta} & 0.87 [0.67, 1.36] & 8.13 & 0.10 & 8.83 [6.02, 12.27] & 2.46 & 3.14 & 0.74 [0.73, 0.77] \\
\textbf{Montenegro} & 0.73 [0.53, 1.17] & 25.45 & 2.18 & 35.62 [23.30, 51.40] & 3.26 & 4.06 & 0.83 [0.73, 1.02] \\
\textbf{Netherlands} & 0.95 [0.73, 1.44] & 10.01 & 2.12 & 11.83 [8.20, 16.20] & 2.14 & 2.86 & 0.79 [0.77, 0.82] \\
\textbf{North Maced.*} & 0.66 [0.45, 1.09] & 25.92 & 0.00 & 38.06 [23.86, 57.46] & 3.52 & 4.14 & 0.84 [0.72, 1.11] \\
\textbf{Norway} & 0.88 [0.68, 1.33] & 1.45 & 0.00 & 1.57 [1.09, 2.13] & 1.97 & 2.68 & 0.76 [0.75, 0.76] \\
\textbf{Poland} & 0.89 [0.67, 1.35] & 19.31 & 11.92 & 31.85 [21.74, 43.94] & 2.76 & 3.56 & 0.85 [0.77, 0.97] \\
\textbf{Portugal} & 1.14 [0.90, 1.67] & 16.40 & 3.49 & 16.20 [11.62, 21.60] & 2.46 & 3.01 & 0.80 [0.77, 0.84] \\
\textbf{Romania} & 0.92 [0.70, 1.40] & 15.69 & 11.65 & 26.55 [18.23, 36.62] & 2.57 & 3.38 & 0.83 [0.76, 0.92] \\
\textbf{Serbia} & 0.96 [0.71, 1.47] & 9.85 & 17.62 & 24.82 [16.90, 34.73] & 2.84 & 3.62 & 0.79 [0.74, 0.88] \\
\textbf{Slovakia} & 0.78 [0.56, 1.23] & 22.39 & 5.02 & 33.02 [21.71, 47.26] & 2.86 & 3.56 & 0.85 [0.77, 1.01] \\
\textbf{Slovenia} & 1.02 [0.79, 1.53] & 20.35 & 1.46 & 20.15 [14.13, 27.30] & 2.44 & 3.13 & 0.81 [0.77, 0.87] \\
\textbf{Spain} & 1.08 [0.87, 1.58] & 17.26 & 4.79 & 18.69 [13.51, 24.65] & 2.49 & 3.11 & 0.80 [0.77, 0.85] \\
\textbf{Sweden} & 1.00 [0.79, 1.48] & 14.16 & 0.00 & 13.40 [9.55, 17.91] & 1.98 & 2.71 & 0.82 [0.79, 0.85] \\
\textbf{Switzerland} & 0.99 [0.78, 1.47] & 12.29 & 0.86 & 12.50 [8.84, 16.75] & 2.18 & 2.88 & 0.79 [0.77, 0.82] \\
\textbf{United Kingdom} & 0.94 [0.75, 1.42] & 22.89 & 1.85 & 24.35 [17.22, 32.65] & 2.31 & 2.98 & 0.85 [0.80, 0.93] \\
\bottomrule
\end{tabular}
\end{table}

\begin{table}
\caption{\textbf{Overview of Pearson and semi-partial correlation coefficients between the estimated prevalence $\alpha$ and variable $Y$.} $Z$ denotes the confounding variable and the relative correlation change is defined by 1 - semi-partial / Pearson and given in percent. Note that a relative correlation change of $<-100\%$ can arise if the Pearson correlation is negative but the semi-partial correlation is positive. Values in brackets denote the 95\% CIs.}\label{sup_tab:correlations}
\centering
\resizebox{\textwidth}{!}{
\begin{tabular}{llllll}
\toprule
Variable $Y$ & Pearson $r$ & Partial ($Z = \momRatio$)  & Rel. change ($Z=\avOutHouseRepNumI$) &  Partial ($Z = \momRatio$) & Rel. Change ($Z=\avOutHouseRepNumI$) \\
\midrule
HDI & -0.79 [-0.80, -0.77] & -0.52 [-0.69, -0.12] & -34\% [-94\%, -11\%] & -0.72 [-0.74, -0.67] & -8\% [-19\%, -4\%]\\
% ()& $\SI{2.15e-8}$ [$\SI{2.15e-8}$, $\SI{2.15e-8}$] & $\SI{1.51e-3}$ [$\SI{1.70e-6}$, $\SI{0.69}$] & - & $\SI{1.09e-6}$ [$\SI{4.18e-7}$, $\SI{2.30e-5}$] &  \\
GNI & -0.64 [-0.66, -0.63] & -0.39 [-0.55, -0.05] & -40\% [-102\%, -14\%] & -0.63 [-0.64, -0.60] & -2\% [-11\%, 2\%] \\
 %()& $\SI{2.90e-5}$ [$\SI{1.85e-5}$, $\SI{4.57e-5}$] & $\SI{0.021}$ [$\SI{0.41e-3}$, $\SI{0.83}$] & - & $\SI{4.18e-5}$ [$\SI{2.82e-5}$, $\SI{0.26e-3}$] &  \\
Life expectancy & -0.69 [-0.69, -0.68] & -0.56 [-0.65, -0.28] & -19\% [-67\%, -35\%] & -0.48 [-0.50, -0.45] & -30\% [-36\%, -27\%] \\
%() & $\SI{5.61e-6}$ [$\SI{5.33e-6}$, $\SI{7.52e-6}$] & $\SI{0.49e-3}$ [$\SI{1.72e-5}$, $\SI{0.20}$] & - & $\SI{0.0034}$ [$\SI{0.0022}$, $\SI{0.0086}$] &  \\
Years of schooling & -0.37 [-0.37, -0.37] & -0.16 [-0.27, 0.06] & -57\% [-126\%, -24\%] & -0.46 [-0.46, -0.45] & +25\% [21\%, 26\%] \\
%() & $\SI{0.030}$ [$\SI{0.029}$, $\SI{0.030}$] & $\SI{0.36}$ [$\SI{0.10}$, $\SI{0.95}$] &-  & $\SI{0.0054}$ [$\SI{0.0053}$, $\SI{0.0074}$] &  \\
\bottomrule
\end{tabular}
}
\end{table}

\begin{table}\caption{\textbf{Statistical significance of results in Tab.~\ref{sup_tab:correlations}.} Values and brackets denote the median and 95\% CI for the $p$ value distribution associated to the different correlations.}
\resizebox{\textwidth}{!}{
\begin{tabular}{llll}
\toprule
Variable $Y$ & $p$ (Pearson $r$) & $p$ (Partial, $Z = \eta^*$) & $p$ (Partial $Z=\bar R_{\rm out}$) \\
\midrule
HDI & $\num{2.15e-8}$ [$\num{2.15e-8}$, $\num{2.15e-8}$] & $\num{1.51e-3}$ [$\num{1.70e-6}$, $\num{0.69}$] & $\num{1.09e-6}$ [$\num{4.18e-7}$, $\num{2.30e-5}$] \\
GNI & $\num{2.90e-5}$ [$\num{1.85e-5}$, $\num{4.57e-5}$] & $\num{0.021}$ [$\num{0.41e-3}$, $\num{0.83}$] & $\num{4.18e-5}$ [$\num{2.82e-5}$, $\num{0.26e-3}$] \\
Life expectancy & $\num{5.61e-6}$ [$\num{5.33e-6}$, $\num{7.52e-6}$] & $\num{0.49e-3}$ [$\num{1.72e-5}$, $\num{0.20}$] & $\num{0.0034}$ [$\num{0.0022}$, $\num{0.0086}$] \\
Years of schooling & $\num{0.030}$ [$\num{0.029}$, $\num{0.030}$] & $\num{0.36}$ [$\num{0.10}$, $\num{0.95}$] & $\num{0.0054}$ [$\num{0.0053}$, $\num{0.0074}$] \\
\bottomrule
\end{tabular}
}
\end{table}

\section{Supplementary Figures}

\begin{figure}
    \centering
    \includegraphics[width=12cm]{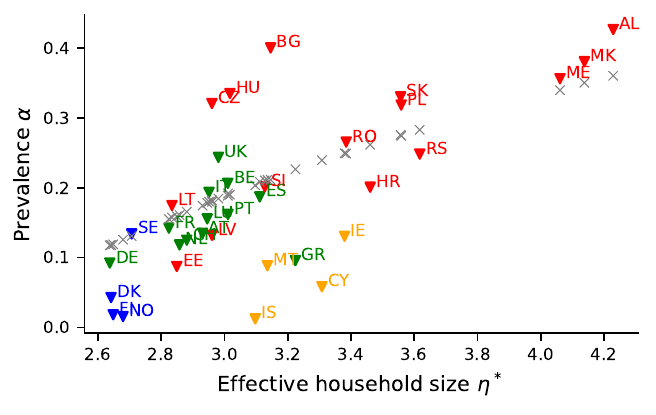}
    \caption{\textbf{Relation between prevalence and effective household sizes.} Same as Fig.~\ref{fig:results_1}A, but with ISO-2 country codes. Theoretical prevalences (gray crosses) are computed assuming a common European out-household reproduction number $R_{\rm out, Europe}$ in all countries. $R_{\rm out, Europe}$ is computed by treating Europe as a proper country. }
    \label{fig:supplementary-mainresults_withlabels}
\end{figure}

\begin{figure}[htbp]
    \centering
    \includegraphics[width=12cm]{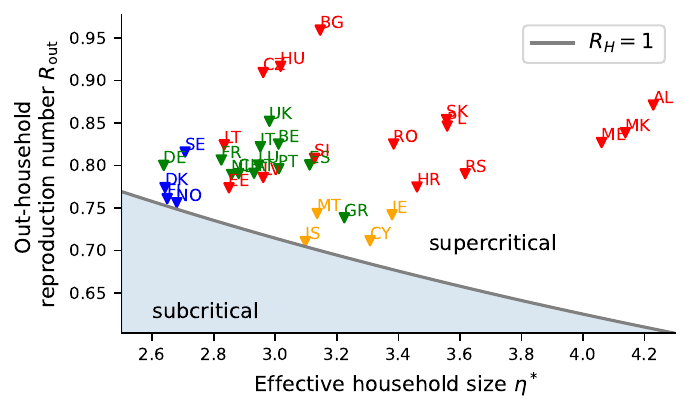}
    \caption{\textbf{Critical out-household reproduction number $\avOutHouseRepNumCrit = \Boost^{-1}$.} The effective household size $\momRatio$ defines a critical out-household reproduction number that, if exceeded, leads to a household reproduction number $\RH > 1$. Countries with a high effective household size $\momRatio$ have to reduce the out-household reproduction number substantially lower to contain an outbreak than countries with a low $\momRatio$. Note that all countries lie in the supercritical region, as they all experienced COVID-19 outbreaks.}
    \label{fig:critical-outhousehold}
\end{figure}